\documentclass{ws-procs9x6}

\def\al{\alpha}
\def\be{\beta}
\def\ga{\gamma}
\def\de{\delta}
\def\ze{\zeta}
\def\ka{\kappa}
\def\la{\lambda}
\def\si{\sigma}
\def\om{\omega}
\def\fr#1#2{{{#1} \over {#2}}}
\def\ket#1{|{#1}\rangle}
\def\half{{\textstyle{1\over 2}}}

\def\tmf{ \widetilde{m}_F }
\def\hmf{ \widehat{m}_F }

\def\pr#1{{#1}^\prime}

\def\tb{\tilde{b}}
\def\td{\tilde{d}}
\def\tg{\tilde{g}}
\def\tc{\tilde{c}}

\def\codt{\cos{\om_s T_s}}
\def\sodt{\sin{\om_s T_s}}
\def\ctodt{\cos{2\om_s T_s}}
\def\stodt{\sin{2\om_s T_s}}

\def\ca{\cos \al}
\def\sa{\sin \al}
\def\cz{\cos \ze}
\def\sz{\sin \ze}

\newcommand{\beq}{\begin{equation}}
\newcommand{\eeq}{\end{equation}}
\newcommand{\bea}{\begin{eqnarray}}
\newcommand{\eea}{\end{eqnarray}}
\newcommand{\rf}[1]{(\ref{#1})}

\begin{document}

\title{Testing Lorentz Symmetry in Space}

\author{Neil Russell}

\address{Physics Department,\\
Northern Michigan University, \\
1401 Presque Isle Avenue, Marquette, MI 49855, USA\\
E-mail: nrussell@nmu.edu}

\maketitle

\abstracts{
{\it Presented at the Third Meeting on CPT and Lorentz Symmetry,
Indiana University, Bloomington IN, 4-6 August, 2004.} \\
\\
Atomic clocks, masers, and other precision oscillators
are likely to be placed on the International Space Station and other
satellites in the future.
These instruments will have the potential
to measure Lorentz-violation coefficients, and in particular may
provide access to parts of the Lorentz-violation coefficient space
at levels not accessible with Earth-based experiments.
The basic issues are outlined
in this proceedings.
}

\section{Lorentz-Violating Standard-Model Extension (SME)}
The Standard-Model Extension (SME) is essentially
the conventional Standard-Model lagrangian of particle physics
plus all possible coordinate-independent Lorentz- and CPT-violating
terms constructed from the conventional fields
of particle physics.\cite{sme,cpt01}
The additional terms could arise in a more fundamental theory,
for example string theory.\cite{strings}
Since the symmetry-violating effects are known to be small,
perturbative methods can be adopted to
calculate the effects in any experimental context.
Calculations or measurements
for the SME in various systems
include investigations of
mesons,\cite{hadrons}
neutrino oscillations,\cite{neutrino}
spin-polarized matter,\cite{spinpol}
hydrogen and antihydrogen,\cite{hbar}
Penning traps,\cite{penning}
muons,\cite{muon}
cosmological birefringence,\cite{birefr}
electromagnetic cavities,\cite{cavities}
electromagnetostatics, \cite{emstatics}
and \u{C}erenkov radiation.\cite{cerenkov}
Various other issues,
including the SME in curved spacetime,\cite{gravity}
have been examined
in the literature.\cite{miscell}

An SME analysis of clock comparison experiments\cite{clak}
provides a comprehensive framework
for relating various tests.\cite{clockcomp}
These experiments search for signals
that are due to rotations and accelerations of the laboratory
relative to an inertial reference frame.
It is therefore natural to consider
clock-comparison experiments performed
in space since the laboratory motion
offers various advantages.
These proceedings provide an overview
of the basic results of this analysis.\cite{space1,space2}

\section{General Clock-comparison Experiments}
An atomic clock is
a device that provides a stable transition frequency
in a particular type of atomic system.
For most atoms of interest,
the total atomic angular momentum and its projection
along the quantization axis are conserved to a high precision,
so the quantum states can be labeled as
$\ket{F, m_F}$.
The shift in the energy levels due to the SME is found
using a perturbation calculation giving
\bea
\de E(F,m_F) &=& \hmf \sum_w (\be_w\tb_3^w + \de_w\td_3^w + \ka_w\tg_d^w)
\nonumber \\
&+& \tmf \sum_w (\ga_w\tc_q^w + \la_w\tg_q^w)
\quad .
\label{AtomicShift}
\eea
The constants
$\hmf$ and $\tmf$
are ratios of Clebsch-Gordan coefficients
given by
\beq
\hmf:= \fr{m_F}{F}
\quad , \qquad
\tmf:= \fr{3m_F^2-F(F+1)}{3F^2-F(F+1)}
\quad .
\label{tildemf}
\eeq
In Eq.\ \rf{AtomicShift}, the five tilde quantities are
specific combinations of the coefficients for Lorentz violation
within the SME.
In the case of $\td_3^w$,
the definition is
\beq
\td_3^w := m_w d_{03}^w +\half m_w d_{30}^w-\half H_{12}^w
 \quad .
\label{bdcgtilde}
\eeq
Similar definitions apply
for the remaining four tilde coefficients.\cite{clak}
Noting that $m_w$ is the mass of particle $w$,
all five tilde coefficients have dimensions of mass.
The index $w$ is to be replaced with $p$ for proton, $e$ for electron,
or $n$ for neutron.
The numerical subscripts refer to
the laboratory-frame coordinate system,
in which the third coordinate is the quantization axis by convention.
Interestingly,
these five tilde combinations are
the only SME parameter combinations that can be bounded
in clock-comparison experiments with ordinary matter.
The aim of this work is to consider ways that
atomic clock transition frequencies may be used to
detect these tilde quantities.
The five Greek-letter coefficients
$\be_w$, $\ga_w$, $\de_w$, $\ka_w$, $\la_w$
appearing in Eq.\ \rf{AtomicShift} are linear
combinations of expectation values
calculated for the state $\ket{F,F}$
of particular operators in the nonrelativistic
hamiltonian for the particle $w$.
For example,
in the case of $\de_w$, the expression is:
\bea
\de_w :=
 \fr 1 {m_w^2}
 \sum_{N=1}^{N_w}\langle [p_3p_j\si^j]_{w,N} \rangle
 \quad ,
\label{bgdkl}
\eea
where $p_j$ are the momentum operators
and $\si^j$ are the three Pauli matrices.
These quantities are calculated for each particle of type $w$
in a specific atom and the index $N$ labels each of the $N_w$ particles
of that type;
for example,
in $^{133}$Cs, $N_p = N_e = 55$ and $N_n = 78$.

To calculate the values of $\de_w$ and the other similar coefficients
would require a detailed understanding of the many-body nuclear physics.
However, reasonable approximations can be made within specific nuclear models.
Dimensional arguments indicate that
$\be_w$ is of order unity,
and the other quantities are suppressed by factors of about
$K_p\approx K_n\simeq 10^{-2}$ and $K_e\simeq 10^{-5}$.

The frequency output $f(B_3)$ of a typical atomic clock is
determined by the difference between two energy levels
and in general depends on the magnetic field
projected on the quantization axis, $B_3$.
Including the Lorentz-violating effects $\de\om$,
the output frequency $\om$
is expressed as
\beq
\om = f(B_3) + \de \om .
\label{FullFrequency}
\eeq
The transition frequency $\om$ is affected by both of the
levels in the transition
$(F,m_F) \rightarrow (\pr{F},\pr{m}_F)$,
so $\de\om$ is determined from
\beq
\de\om = \de E(F,m_F)- \de E(\pr{F},\pr{m}_F) .
\label{FrequencyComparison}
\eeq

\section{Standard Inertial Reference Frame}
The Lorentz-violating effects in equation \rf{AtomicShift}
are contained in the SME tilde quantities,
which are tensors under observer transformations.
Thus,
their components in one inertial reference frame
are related to those in another
by the corresponding rotation or boost between observers.
However,
unlike the energy-momentum tensor,
for example,
they are not integrated from controllable experimental source configurations.
They are instead fixed in space.
In conventional physics,
results are independent of the orientation or velocity of the laboratory,
but this is no longer true
since the interaction of the experiment
with this fixed Lorentz-violating background
introduces time-dependent effects.
A measurement of $\td_3^w$,
for example,
is time-dependent
since the third component in the laboratory frame
is changing its orientation as the Earth rotates.

The time dependence is determined by the laboratory motion
relative to a standard reference frame.
By convention,
this frame is centered on the Sun
with $Z$ axis  parallel to the rotation axis of the Earth,
and with $X$ axis pointing at the vernal equinox on the celestial sphere.
The time $T$ is measured from the vernal equinox in the year 2000.
Measurement of the SME coefficients in the standard frame
is done using the laboratory trajectory
through a sequence of linear transformations.
For the case of a satellite,
the motion is a combination of the
circular motion of the Earth around the Sun
and the circular motion of the satellite around the Earth.
As an example of one of the laboratory-frame quantities
expressed in terms of the inertial frame,
the expression for $\td_3$ is
\bea \td_3 &=&
\codt \Big\{ \Big[
 \td_X(-\sa\cz) + \td_Y(\ca\cz) + \td_Z(\sz)
 \Big]
 \nonumber \\
&& \phantom{--------.} +\be_\oplus\Big[
 {\mbox{seasonal Sun-frame tilde terms}}
 \Big] \Big\}
 \nonumber \\
&+& \sodt \Big\{ \Big[
 \td_X( -\ca ) + \td_Y( -\sa )
 \Big]
 \nonumber \\
&& \phantom{--------.} +\be_\oplus\Big[
 {\mbox{seasonal Sun-frame tilde terms}}
 \Big] \Big\}
 \nonumber \\
&+& \ctodt \Big\{
 \be_s\Big[
 {\mbox{constant Sun-frame tilde terms}}
 \Big] \Big\}
 \nonumber \\
&+& \stodt \Big\{
 \be_s\Big[
 {\mbox{constant Sun-frame tilde terms}}
 \Big] \Big\}
 \nonumber \\
&& \phantom{---} + \Big\{
 \be_s\Big[
 {\mbox{constant Sun-frame tilde terms}}
 \Big] \Big\} .
\label{Explicitd3}
\eea
Here,
the $z$ or $3$ direction in the lab is oriented along
the velocity vector of the satellite relative to the Earth,
while the $x$ direction points towards the center of the Earth.
The satellite time $T_s$
is related to the Sun-based time
by $T=T_s + T_0,$
where $T=T_0$ is the time of a selected
ascending node of the satellite.
Other satellite orbital elements in the expression are
the right ascension $\al$ of the ascending node
and the inclination $\ze$ between the orbital axis and the Earth's axis.
For the International Space Station,
$\om_s \approx 2\pi/92 \mbox{min}$
and $\be_s \approx 3 \times 10^{-5}$
are the orbital frequency and speed
relative to the Earth.
The speed of the Earth is $\be_\oplus \approx 1.0 \times 10^{-4}$,
and the seasonal terms refer to cyclic variations
with angular frequency $2\pi/(\mbox{one sidereal year})$.

In Eq.~\rf{Explicitd3},
only the sun-frame tilde components
$\td_X$, $\td_Y$, and $\td_Z$ appear explicitly.
Others appear in the seasonal and constant Sun-frame expressions,
which are given in full and in tabular form in Ref.~\refcite{space2}.
Both single and double frequencies appear in the expressions,
and can be understood as arising from single-
and double-index coefficients in the SME.
An advantage of using a satellite is
the relatively high frequency $\om_s$
which reduces the limitation of clock stability over time.
Use of a turntable in a ground-based laboratory,
as is being done in some experiments,
offers a similar stability payoff although the velocity factor
$\be_s$ is reduced 16-fold to the value
$\be_L \approx 1.6 \times 10^{-6}$.

The coefficients that a particular clock-comparison
experiment could detect in principle
depend on the atoms of the clock
and the transition used.
An analysis has been done for rubidium clocks,
cesium clocks, and hydrogen masers.\cite{space2}
Similar techniques can be applied to other systems.

\section{Discussion}
There are 120 coefficients
that in principle clock-comparison experiments
can detect at leading order,
consisting of 40 for each of the three basic sub-atomic particles.
About half of these coefficients are
suppressed by a factor of $\be_s$,
indicating that detection
of these coefficients may be enhanced
in a satellite moving at high $\be_s$.
A number of coefficients have been probed with earth-based experiments,
even though the lab speed relative to the Earth is
an order of magnitude less than in orbit.
If experiments were done today with cesium and rubidium atomic clocks
in space, several dozen unmeasured coefficients would be accessed.
Others would be accessible with different clocks,
and in principle, all 120 coefficients are accessible from space-based
clock-comparison experiments.

\end{document}